\documentclass[11pt]{article}

\usepackage{troffms}

\begin{document}

\footer{\rm\thepage}

\title{Two simple solutions  \\
of nonlinear Fokker - Planck equation \\
for incompressible fluid \\
}
\author{Igor A. Tanski \\\
       Moscow, Russia \\
       tanski.igor.arxiv@gmail.com \\
}
        
\rule{6in}{1pt}
        
\begin{abstract}
In this article we derive two simple solutions of nonlinear Fokker - Planck equation for incompressible fluid and investigate their properties
\end{abstract}

\shead{Keywords}

\par\noindent
Fokker-Planck equation, continuum mechanics, incompressible fluid

\rule{6in}{1pt}

\section{Introduction}

\par
In our previous work [1] we found following nonlinear Fokker - Planck equation for incompressible fluid:

$$
\rho =
\int_V n\  dv_1 dv_2 dv_3 = const ,
\eqno (1)$$

$$
{\partial n  \over \partial t} +
v_k {\partial n  \over \partial x_k} - 
\alpha\  { \partial (v_j n)  \over \partial v_j} -
{1 \over \rho }{\partial n  \over \partial v_j}
{\partial p  \over \partial x_j} 
= k\  {\partial^2 n  \over \partial v_j \partial v_j} .
\eqno (2)$$

\par\noindent
where
\par\noindent
$n = n(t, x_1 , x_2 , x_3 , v_1 , v_2 , v_3 )$ - density;
\par\noindent
$p$ - pressure;
\par\noindent
$t$ - time variable;
\par\noindent
$x_1 , x_2 , x_3 $ - space coordinates;
\par\noindent
$v_1 , v_2 , v_3$ - velocities;
\par\noindent
$\alpha$ - coefficient of damping;
\par\noindent
$k$ - coefficient of diffusion.

\par
In this work we constructed only stationary solution of the system (1-2) with zero average Maxwell velocities distribution and Pascal pressure field. In the present work we construct some another simple solutions, which generalize previous results.

\section{Zero pressure solutions of nonlinear Fokker - Planck equation for incompressible fluid}

\par
First of all we consider zero pressure field

$$
p = 0 . 
\eqno (3)$$

\par
In this case

$$
{\partial n  \over \partial t} +
v_k {\partial n  \over \partial x_k} - 
\alpha\  { \partial (v_j n)  \over \partial v_j} 
= k\  {\partial^2 n  \over \partial v_j \partial v_j} .
\eqno (4)$$

\par
This is simply standard Fokker - Planck equation, but solutions must satisfy additional equation (1). We know, that solution of equation (4) has following form (see [2]):

$$
n(t, x_j , v_j ) =
\sum_{ m_1 = - \infty }^{ + \infty }
\sum_{ m_2 = - \infty }^{ + \infty }
\sum_{ m_3 = - \infty }^{ + \infty }
\sum_{ n_1 = 0 }^{ \infty }
\sum_{ n_2 = 0 }^{ \infty }
\sum_{ n_3 = 0 }^{ \infty }
\exp \left[
- t\ 
\left( 
\alpha \  \sum_{{j=1}}^3 n_j +
k\  \left(
{2 \pi   \over \alpha }\right)^2
\sum_{{j=1}}^3
\left(
{ m_j   \over  a_j }
\right)^2 
\right) 
\right]
\times
\eqno (5)$$
$$
\times
A_{ m_1 m_2 m_3 n_1 n_2 n_3 } \phi_{ m_1 m_2 m_3 n_1 n_2 n_3 } ,
$$

\par\noindent
where

$$
\phi_{ m_1 m_2 m_3 n_1 n_2 n_3 } =
\prod_{{j=1}}^{{j=3}}
\exp 
\left( 
2 \pi i {m_j \over a_j }( x_j - {v_j \over \alpha }) 
\right)
\exp
\left(
- {\alpha \over 2k }\ 
v_j^2
\right)
H_{{n}_j}
\left(
\sqrt {{\alpha \over 2k }}
\left(
v_j +
{4 \pi i m_j k  \over \alpha^2 a_j}
\right)
\right)
 .
\eqno (6)$$

\par
Comparing first term of (5) with constant $\rho$ in the LHS of (1), we find:

$$
A_{ m_1 m_2 m_3 0 0 0 } = \rho\  \left( {\alpha \over 2 \pi k} \right)^{3/2}  .
\eqno (7)$$

\par
(1) implies obviously, that for all another terms with $n_1 = n_2 = n_3 = 0$ coefficient is zero.

$$
A_{ m_1 m_2 m_3 0 0 0 } = 0\ \ \ m_1 \ge0\  or\  m_2 \ge0\  or\  m_3 \ge0 ;
\eqno (8)$$

\par
For all another $n_i$ we group terms with the same set of indices $m_i$ and equal sums of indices $n_i$:

$$
G_s ( m_1 , m_2 , m_3 ) = 
\sum_{{n}_1 + n_2 + n_3 = s}
\exp \left[
- t\ 
\left( 
\alpha \  \sum_{{j=1}}^3 n_j +
k\  \left(
{2 \pi   \over \alpha }\right)^2
\sum_{{j=1}}^3
\left(
{ m_j   \over  a_j }
\right)^2 
\right) 
\right]
\times
\eqno (9)$$
$$
\times
A_{ m_1 m_2 m_3 n_1 n_2 n_3 } \phi_{ m_1 m_2 m_3 n_1 n_2 n_3 } ,
$$

\par\noindent
where group index 

$$
s = \sum_{{j=1}}^3 n_j ;
\eqno (10)$$

\par
We considered such groups in our works [3] ($s=0$) and [4] ($s=1$).

\par
For each group $G_s$ amplitude decreases with time as $\exp (- \alpha s t )$. Therefore we can carry out time dependent multiplier before the sum:

$$
G_s ( m_1 , m_2 , m_3 ) = 
\exp \left[
- t\ 
\left( 
\alpha \  \sum_{{j=1}}^3 n_j +
k\  \left(
{2 \pi   \over \alpha }\right)^2
\sum_{{j=1}}^3
\left(
{ m_j   \over  a_j }
\right)^2 
\right) 
\right]
\times
\eqno (11)$$
$$
\times
\sum_{{n}_1 + n_2 + n_3 = s}
A_{ m_1 m_2 m_3 n_1 n_2 n_3 } 
\phi_{ m_1 m_2 m_3 n_1 n_2 n_3 } ,
$$

\par
Let us denote the integral:

$$
I_{{m}_j n_j} = 
\int_{{-} \infty}^{\infty}
\exp 
\left( 
- 2 \pi i { {m_j \over a_j} {v_j \over \alpha} }\right)
\exp
\left(
- {\alpha \over 2k }\ 
v_j^2
\right)
H_{{n}_j}
\left(
\sqrt {{\alpha \over 2k }}
\left(
v_j +
{4 \pi i m_j k  \over \alpha^2 a_j}
\right)
\right)\ 
d v_j .
\eqno (12)$$

\par
Equation (1) gives following orthogonality condition for each group:

$$
\sum_{{n}_1 + n_2 + n_3 = s}
A_{ m_1 m_2 m_3 n_1 n_2 n_3 } 
I_{{m}_1 n_1} 
I_{{m}_2 n_2}
I_{{m}_3 n_3} = 0\ \ \ \ s > 0 ;
\eqno (13)$$

\par
To calculate the integral (1) we perform following substitutions:

$$
- {\alpha \over 2k }\left(
v^2 + {4 \pi i m k v  \over \alpha^2 a}
\right) =
- {\alpha \over 2k }\left[
\left(
v + {2 \pi i m k   \over \alpha^2 a}
\right)^2 +
\left(
{2 \pi m k   \over \alpha^2 a}
\right)^2
\right] .
\eqno (14)$$

$$
\xi =
\sqrt {{\alpha \over 2k }}
\left(
v + {2 \pi i m k   \over \alpha^2 a}
\right) .
\eqno (15)$$

\par
This gives:

$$
I_{mn} =
\sqrt {{2k \over \alpha }}
\exp \left[
-
{\alpha \over 2k }\left(
{2 \pi m k  \over \alpha^2 a}
\right)^2
\right] \ 
\int_{{-} \infty}^{\infty}
\exp \left[
- \xi^2
\right]\ 
H_n
\left(
\xi + 
\sqrt {{\alpha \over 2k }}
{2 \pi i m k  \over \alpha^2 a}
\right)
d \xi
.
\eqno (16)$$

\par
The addition theorem for Hermit polynomials is (see [5]):

$$
H_n ( z_1 + z_2 ) =
2^{{-n} / 2}
\sum_{k=0}^n
\left(
\matrix {n \cr k}
\right)
H_k (z_1 \sqrt 2 )
H_{n-k} (z_2 \sqrt 2 ) .
\eqno (17)$$

\par
Orthogonality condition for Hermit polynomials is:

$$
\int_{{-} \infty}^{\infty}
\exp 
\left(
{- x^2}
\right)
H_m (x)
H_n (x)
dx =
\delta_{mn}
\sqrt \pi 2^n n! .
\eqno (18)$$

\par
Apply (17) to (16)

$$
H_n \left( \xi + 
\sqrt {{\alpha \over 2k }}
{2 \pi i m k  \over \alpha^2 a} \right) =
2^{{-n} / 2}
\sum_{k=0}^n
\left(
\matrix {n \cr k}
\right)
H_k ( \xi \sqrt 2 )
H_{n-k} 
\left(
\sqrt {{\alpha \over k }}
{2 \pi i m k  \over \alpha^2 a}
\right) =
\eqno (19)$$
$$
=
2^{{-n} / 2}
H_0 ( \xi \sqrt 2 )
H_n 
\left(
\sqrt {{\alpha \over k }}
{2 \pi i m k  \over \alpha^2 a}
\right) + 
2^{{-n} / 2}
n
H_1 ( \xi \sqrt 2 )
H_{n-1} 
\left(
\sqrt {{\alpha \over k }}
{2 \pi i m k  \over \alpha^2 a}
\right) +
...
.
$$

\par
In the first version of this article we applied (18) to (19) an got erroneous result. In fact we can not apply (18) to (19), because argument of $H_n$ in (19) is $( \xi \sqrt 2 )$, not $( \xi )$ - as in (18). So addition theorem in the form (17) is useless for integral $I_{mn}$ calculation.

\par
Instead of this there are three correct ways to obtain integral $I_{mn}$ value:
\par\noindent
- direct use of generating function of Hermite polynomials. We consider this way in APPENDIX 1.
\par\noindent
- another form of addition theorem. We consider this way in APPENDIX 2.
\par\noindent
- use of Fourier transform. We consider this way in APPENDIX 3.

\par
This last way we used in our work [2]. This makes our error in the first version of this paper even more inexcusable.

\par
The result is:

$$
I_{mn} =
\sqrt {{2 \pi k  \over a }}
\left(
{2k \over \alpha }\right)^{n/2}
\exp \left[
- {k \over 2 \alpha}
\left(
{2 \pi m  \over \alpha a}
\right)^2
\right]\ 
\left(
{2 \pi i m  \over \alpha a}
\right)^n .
\eqno (20)$$

\par
Carry out all multipliers, which depend only on $m_i$ or $s = \sum n_i$. This gives following orthogonality condition for the group:

$$
\sum_{{n}_1 + n_2 + n_3 = s}
A_{ m_1 m_2 m_3 n_1 n_2 n_3 } 
\left(
{m_1  \over a_1}
\right)^{{n}_1}
\left(
{m_2  \over a_2}
\right)^{{n}_2}
\left(
{m_3  \over a_3}
\right)^{{n}_3} = 0\ \ \ \ s > 0 .
\eqno (21)$$

\par
For example for the group $s=1$ we have following condition:

$$
A_{ m_1 m_2 m_3 1 0 0 } 
{m_1  \over a_1} +
A_{ m_1 m_2 m_3 0 1 0 } 
{m_2  \over a_2} +
A_{ m_1 m_2 m_3 0 0 1 } 
{m_3  \over a_3} 
 = 0\ \ \ \ s=1 .
\eqno (22)$$

\par
This condition means physically, that divergence of the average velocity field is zero. When initial velocity field satisfy the set of such conditions, it can be prolonged for all times $t0$ with zero pressure field.

\section{Average velocities}

\par
It is interesting to derive expressions for average velocities. For this purpose we denote:

$$
J_{{m}_j n_j} = 
\int_{{-} \infty}^{\infty}
\exp 
\left( 
- 2 \pi i { {m_j \over a_j} {v_j \over \alpha} }\right)
\exp
\left(
- {\alpha \over 2k }\ 
v_j^2
\right)
H_{{n}_j}
\left(
\sqrt {{\alpha \over 2k }}
\left(
v_j +
{4 \pi i m_j k  \over \alpha^2 a_j}
\right)
\right)\ 
v_j
d v_j .
\eqno (23)$$

\par
In the first version of this paper we tried to calculate $J_{{mn}}$ value using addition theorem (17). But, as before, (17) is inconvenient for this purpose. The correct way of $J_{{mn}}$ value deduction is given in APPENDIX 4.

\par
So we have

$$
J_{{m}_j n_j} =
\sqrt {{2 \pi k  \over \alpha }}
\left(
{2k \over \alpha }\right)^{{n}_j / 2}
\exp \left[
- {k \over 2 \alpha}
\left(
{2 \pi m_j  \over \alpha a_j}
\right)^2
\right]\ 
\left[
-
{k \over \alpha} 
\left( 
{ 2 \pi i m_j   \over \alpha a_j}
\right)^{{n}_j + 1} + 
n_j \ 
\left( 
{ 2 \pi i m_j   \over \alpha a_j}
\right)^{{n}_j - 1}
\right]
.
\eqno (24)$$

$$
J_{{m}_j 0} =
-
\sqrt {{2 \pi k  \over \alpha }}
\exp \left[
- {k \over 2 \alpha}
\left(
{2 \pi m_j  \over \alpha a_j}
\right)^2
\right]\ 
\left[
{k \over \alpha} 
\left( 
{ 2 \pi i m_j   \over \alpha a_j}
\right)
\right]
.
\eqno (25)$$

\par
In this way we get final expressions for $\rho u_i$

$$
\rho u_1 =
\exp \left[
- t\ 
\left( 
\alpha \  \sum_{{j=1}}^3 n_j +
k\  \left(
{2 \pi   \over \alpha }\right)^2
\sum_{{j=1}}^3
\left(
{ m_j   \over  a_j }
\right)^2 
\right) 
\right]
\prod_{{j=1}}^{{j=3}}
\exp 
\left( 
2 \pi i {m_j \over a_j }x_j 
\right)
\times
\eqno (26)$$
$$
\times
\sum_{{n}_1 + n_2 + n_3 = s}
A_{ m_1 m_2 m_3 n_1 n_2 n_3 } 
J_{{m}_1 n_1} 
I_{{m}_2 n_2}
I_{{m}_3 n_3} .
$$

$$
\rho u_2 =
\exp \left[
- t\ 
\left( 
\alpha \  \sum_{{j=1}}^3 n_j +
k\  \left(
{2 \pi   \over \alpha }\right)^2
\sum_{{j=1}}^3
\left(
{ m_j   \over  a_j }
\right)^2 
\right) 
\right]
\prod_{{j=1}}^{{j=3}}
\exp 
\left( 
2 \pi i {m_j \over a_j }x_j 
\right)
\times
\eqno (27)$$
$$
\times
\sum_{{n}_1 + n_2 + n_3 = s}
A_{ m_1 m_2 m_3 n_1 n_2 n_3 } 
I_{{m}_1 n_1} 
J_{{m}_2 n_2}
I_{{m}_3 n_3} .
$$

$$
\rho u_3 =
\exp \left[
- t\ 
\left( 
\alpha \  \sum_{{j=1}}^3 n_j +
k\  \left(
{2 \pi   \over \alpha }\right)^2
\sum_{{j=1}}^3
\left(
{ m_j   \over  a_j }
\right)^2 
\right) 
\right]
\prod_{{j=1}}^{{j=3}}
\exp 
\left( 
2 \pi i {m_j \over a_j }x_j 
\right)
\times
\eqno (28)$$
$$
\times
\sum_{{n}_1 + n_2 + n_3 = s}
A_{ m_1 m_2 m_3 n_1 n_2 n_3 } 
I_{{m}_1 n_1} 
I_{{m}_2 n_2}
J_{{m}_3 n_3} .
$$

\par
In the next section we consider solution

$$
A_{ 0 1 0 1 0 0 } = 1;\ \ 
A_{ 0 1 0 0 1 0 } = 0;\ \ 
A_{ 0 1 0 0 0 1 } = 0.
\eqno (29)$$

\par
In this case

$$
n = 
\rho\  \left( {\alpha \over 2 \pi k} \right)^{3/2} 
\exp \left[
- {\alpha \over 2k }v_j v_j
\right] +
\eqno (30)$$
$$
+
\exp \left[
- t\ 
\left( 
\alpha +
k\  \left(
{2 \pi   \over \alpha a_2 }
\right)^2
\right)
\right]
\exp 
\left( 
{2 \pi i  \over a_2 }( x_2 - {v_2 \over \alpha }) 
\right)
\exp
\left(
- {\alpha \over 2k }\ 
( v_1^2 +
v_2^2 +
v_3^2 )
\right)\ 
H_1
\left(
\sqrt {{\alpha \over 2k }}
v_1
\right) .
$$

\par
Coefficients are equal to

$$
J_{01} =
\sqrt {\pi}
{2 k  \over \alpha },\ \ 
I_{10} =
\sqrt {{2 \pi k  \over \alpha }}
\exp \left[
-
{\alpha \over 2k }\left(
{2 \pi k  \over \alpha^2 a_2}
\right)^2
\right] , \ 
I_{00} = \sqrt {{2 \pi k  \over \alpha }} .
\eqno (31)$$

$$
\rho u_1 = 
{1 \over \sqrt \pi}
\left(
{2 \pi k  \over \alpha}
\right)^2
\exp \left[
-
{\alpha \over 2k }\left(
{2 \pi k  \over \alpha^2 a_2}
\right)^2
\right]\ 
\exp \left[
- t\ 
\left( 
\alpha +
k\  \left(
{2 \pi   \over \alpha a_2 }
\right)^2
\right)
\right]
\exp 
\left( 
{2 \pi i  \over a_2 }x_2 
\right)\ 
 .
\eqno (32)$$

$$
u_2 = u_3 = 0 .
\eqno (33)$$

\par
This is time dependent shift along X-axis with amplitude depending on Y space coordinate. Note, that $u_1$ is periodic on $x_2$ and average $u_1$ on $x_2$ is zero.

\section{Constant pressure gradient solutions of nonlinear Fokker - Planck equation for incompressible fluid}

\par
The case of constant pressure gradient is also very simple. Let us denote

$$
{\partial p  \over \partial x_1} = \alpha h ;\ \ 
{\partial p  \over \partial x_2} = 0;\ \ 
{\partial p  \over \partial x_3} = 0 .
\eqno (34)$$

\par
Equation (2) take the form

$$
{\partial n  \over \partial t} +
v_k {\partial n  \over \partial x_k} - 
\alpha\  { \partial (v_k n)  \over \partial v_k} -
h
{\partial n  \over \partial v_1}
= k\  {\partial^2 n  \over \partial v_k \partial v_k} .
\eqno (35)$$

\par
Perform in (35) the substitution

$$
v_1 ' = v_1 - h ;\ \ v_1 = v_1 ' + h ; \ \ x_1 ' = x_1 - h t;\ \ x_1 = x_1 ' + h t .
\eqno (36)$$

\par\noindent
and (35) transforms to zero pressure case (4).

\par
This follows also from eq. (68) of [1]. More specifically, we use the symmetry group

$$
\xi_3 =
( \alpha f_1 ' - f_1 '' ) x {\partial   \over \partial p} +
f_1 ' (t) {\partial   \over \partial u} +
f_1 (t) {\partial   \over \partial x} ;
\eqno (37)$$

\par\noindent
where

$$
f_1 = t;\ \ f_1 = 1;\ \ f_1 '' = 0 ,
\eqno (38)$$

\par\noindent
that is

$$
v_3 =
\alpha x {\partial   \over \partial p} +
 {\partial   \over \partial u} +
t {\partial   \over \partial x} .
\eqno (39)$$

\par
The invariant solutions are

$$
n (t, x, y, z, u, v, w) = n_0 
(t, x + ht, y, z, u + h , v, w) .
\eqno (40)$$

\par\noindent
where $n_0$ is solution for zero pressure case.

$$
p (t, x, y, z) = \alpha h\  x .
\eqno (41)$$

\par
For the solution (29-30), which does not depend on X coordinate, we get in this way Poiseuille type solution.

\par
In this case

$$
n = 
\exp \left[
- t\ 
\left( 
\alpha +
k\  \left(
{2 \pi   \over \alpha a_2 }
\right)^2
\right)
\right]
\exp 
\left( 
{2 \pi i  \over a_2 }( x_2 - {v_2 \over \alpha }) 
\right)
\times
\eqno (42)$$
$$
\times
\exp
\left(
- {\alpha \over 2k }\ 
( ( v_1 + h )^2 +
v_2^2 +
v_3^2 )
\right)\ 
H_1
\left(
\sqrt {{\alpha \over 2k }}
( v_1 + h )
\right) ;
$$

$$
\rho u_1 = 
- \rho h +
{1 \over \sqrt \pi}
\left(
{2 \pi k  \over \alpha}
\right)^2
\exp \left[
-
{\alpha \over 2k }\left(
{2 \pi k  \over \alpha^2 a_2}
\right)^2
\right]\ 
\exp \left[
- t\ 
\left( 
\alpha +
k\  \left(
{2 \pi   \over \alpha a_2 }
\right)^2
\right)
\right]
\exp 
\left( 
{2 \pi i  \over a_2 }x_2 
\right)\ 
 .
\eqno (43)$$

$$
u_2 = u_3 = 0 .
\eqno (44)$$

\par
We see, that constant increasing of $p$ along X axis results in adding constant velocity in the opposite direction - quite as for Poiseuille flow.

\shead{DISCUSSION}
\par\noindent
In this article we derive two simple solutions of nonlinear Fokker - Planck equation for incompressible fluid and investigate their properties. They are: flows with zero pressure and flows with constant pressure gradient. This short list of solution will be useful for further investigations.

\rule{2in}{1pt}
\shead{REFERENCES}

\begin{IPlist}
\IPitem{{[1]}}
Igor A. Tanski. Fokker - Planck equation for incompressible fluid. 
arXiv:0812.2303v1 [nlin.CD]

\IPitem{{[2]}}
Igor A. Tanski. Spectral decomposition of 3D Fokker - Planck differential operator. arXiv:nlin/0607050v3 [nlin.CD] v3 25 Jun 2007

\IPitem{{[3]}}
Igor A. Tanski. Spectral decomposition approach to macroscopic parameters of Fokker-Planck flows: Part 1. arXiv:0707.3306 v3 17 Sep 2007

\IPitem{{[4]}}
Igor A. Tanski. Spectral decomposition approach to macroscopic parameters of Fokker-Planck flows: Part 2. arXiv:0708.0700 v2 17 Sep 2007

\IPitem{{[5]}}
E. Janke, F. Emde, F. L\"{o}sch. Tafeln h\"{o}herer funktionen: sechste auflage. B. G. Teubner Verlagsgesellschaft, Stuttgart, 1960

\IPitem{{[6]}}
J. Kamp'e' de Feriet. Fonctions de la physique math'e'matique. Centre National de la Recherche Scientifique, 1957

\end{IPlist}\shead{APPENDIX 1}

\par
We start from generating function of Hermite polynomials (see for example [5]):

$$
\sum_{n=0}^{\infty}
H_n (z) {t^n \over n }! =
\exp \left(
- t^2 + 2 z t
\right)
.
\eqno (AP1-1)$$

\par
Substitute $z= z_1 + z_2$

$$
\sum_{n=0}^{\infty}
H_n (z_1 + z_2 ) {t^n \over n }! =
\exp \left(
- t^2 + 2 ( z_1 + z_2 ) t
\right)
.
\eqno (AP1-2)$$

\par
Multiply both sides by $e^{{-} z_1^2}$ and integrate:

$$
\int_{{-} \infty}^{\infty}
e^{{-} z_1^2}
\sum_{n=0}^{\infty}
H_n (z_1 + z_2 ) {t^n \over n }! 
d z_1 =
\int_{{-} \infty}^{\infty}
e^{{-} z_1^2}
\exp \left(
- t^2 + 2 ( z_1 + z_2 ) t
\right)
d z_1 =
\eqno (AP1-3)$$
$$
=
\exp \left(
 2 z_2 t
\right)
\int_{ - \infty}^{\infty}
e^{- ( z_1 - t )^2}
d z_1 =
\sqrt \pi
\exp \left(
 2 z_2 t
\right) =
\sqrt \pi
\sum_{n=0}^{\infty}
{
{\left(
2 z_2 t
\right)^n } \over n! }
.
$$

\par\noindent
(we used Poisson's integral value). Compare coefficients by $t^n$ and get

\boxit{
$$
\int_{{-} \infty}^{\infty}
e^{{-} z_1^2}
H_n (z_1 + z_2 )
d z_1 = 
\sqrt \pi
\left(
2 z_2
\right)^n .
\eqno (AP1-4)$$
}

\par
To get value of $I_{mn}$ integral we substitute $z_2 = \sqrt {{\alpha \over 2k }}
{2 \pi i m k  \over \alpha^2 a} $. We have by definition

$$
I_{mn} =
\sqrt {{2k \over \alpha }}
\exp \left[
-
{\alpha \over 2k }\left(
{2 \pi m k  \over \alpha^2 a}
\right)^2
\right] \ 
\int_{{-} \infty}^{\infty}
\exp \left[
- \xi^2
\right]\ 
H_n
\left(
\xi + 
\sqrt {{\alpha \over 2k }}
{2 \pi i m k  \over \alpha^2 a}
\right)
d \xi
.
\eqno (AP1-5)$$

\par
According (AP1-4)

$$
I_{mn} =
\sqrt {{2 \pi k   \over \alpha }}
\exp \left[
-
{\alpha \over 2k }\left(
{2 \pi m k  \over \alpha^2 a}
\right)^2
\right] \ 
\left(
{2 \alpha  \over k }\right)^{n/2}
\left(
{2 \pi i m k  \over \alpha^2 a}
\right)^n
.
\eqno (AP1-6)$$

\par
So we have

$$
I_{mn} =
2 \pi 
\sqrt {{k \over 2 \pi a}}
(-i)^n
\left(
{2k \over \alpha }\right)^{n/2}
\exp \left[
- {k \over 2 \alpha}
\left(
{2 \pi m  \over \alpha a}
\right)^2
\right]\ 
\left(
- {2 \pi m  \over \alpha a}
\right)^n .
\eqno (AP1-7)$$

$$
I_{mn} =
\sqrt {{2 \pi k  \over a }}
\left(
{2k \over \alpha }\right)^{n/2}
\exp \left[
- {k \over 2 \alpha}
\left(
{2 \pi m  \over \alpha a}
\right)^2
\right]\ 
\left(
{2 \pi i m  \over \alpha a}
\right)^n .
\eqno (AP1-8)$$

\shead{APPENDIX 2}

\par
Use once again (AP1-2). This time we multiply both sides by $e^{{-} z_1^2} H_m ( z_1 )$ and integrate:

$$
\int_{{-} \infty}^{\infty}
e^{{-} z_1^2}
H_m ( z_1 )
\sum_{n=0}^{\infty}
H_n (z_1 + z_2 ) {t^n \over n }! 
d z_1 =
\int_{{-} \infty}^{\infty}
e^{{-} z_1^2}
H_m ( z_1 )
\exp \left(
- t^2 + 2 ( z_1 + z_2 ) t
\right)
d z_1 =
\eqno (AP2-1)$$
$$
=
\exp \left(
 2 z_2 t
\right)
\int_{{-} \infty}^{\infty}
H_m ( z_1 )
e^{{-} ( z_1 - t )^2}
d z_1 =
\exp \left(
 2 z_2 t
\right)
\int_{{-} \infty}^{\infty}
H_m ( x + t )
e^{{-} x^2}
dx =
$$
$$
=
\exp \left(
 2 z_2 t
\right) 
\sqrt \pi
\left(
2 t
\right)^m
=
\sqrt \pi
\left(
2 t
\right)^m
\sum_{k=0}^{\infty}
{
{\left(
2 z_2 t
\right)^k } \over k! }
.
$$

\par
Compare coefficients by $t^n$

\boxit{
$$
\int_{{-} \infty}^{\infty}
e^{{-} t^2}
H_m (t)
H_n (t + x )
d t = 
\left\{
\matrix {
0\ \ \ n < m 
\cr 
\sqrt \pi 2^n x^{n-m} {n!} / {(n-m)!}\ \ \ n \ge m 
}
\right. .
\eqno (AP2-2)$$
}

\par
This integral gives coefficients of following Fourier decomposition:

\boxit{
$$
H_n ( t + x ) =
\sum_{k=0}^n
\left(
\matrix {n \cr k}
\right)\ 
(2 x )^{n-k}
H_k (t) 
\eqno (AP2-3)$$
}

\par
This is another form of addition theorem (compare with (17)), which we can use to calculate $I_{mn} $ integral value.

\par
Another way to deduct (AP2-3), which considers (AP2-3) as Taylor series, is given in [6].

\shead{APPENDIX 3}

\par
By definition

$$
I_{mn} = 
\int_{{-} \infty}^{\infty}
 \phi_{mn} dv .
\eqno (AP3-1)$$

\par
According to [2] Fourier transform of $\phi_{mn}$ is

$$
\phi_{mn} ( v ) \to
\sqrt {{k \over 2 \pi a}}
(-i)^n
\left(
{2k \over \alpha }\right)^{n/2}
\exp \left[
- {k \over 2 \alpha}
\left(
q + {2 \pi m  \over \alpha a}
\right)^2
\right]\ 
\left(
q - {2 \pi m  \over \alpha a}
\right)^n 
=
F_{mn} (q)
.
\eqno (AP3-2)$$

\par
According to definition of Fourier transformation we could express the value of integral of any function through the value of its transform in zero point:

$$
I_{mn} = 
2 \pi F_{mn} (0) =
2 \pi
\times
\eqno (AP3-3)$$
$$
\times
\left[
\sqrt {{k \over 2 \pi a}}
(-i)^n
\left(
{2k \over \alpha }\right)^{n/2}
\exp \left[
- {k \over 2 \alpha}
\left(
q + {2 \pi m  \over \alpha a}
\right)^2
\right]\ 
\left(
q - {2 \pi m  \over \alpha a}
\right)^n
\right] \ \ \ q=0
 .
$$

\par
This gives

$$
I_{mn} = 
2 \pi 
\sqrt {{k \over 2 \pi a}}
(-i)^n
\left(
{2k \over \alpha }\right)^{n/2}
\exp \left[
- {k \over 2 \alpha}
\left(
{2 \pi m  \over \alpha a}
\right)^2
\right]\ 
\left(
- {2 \pi m  \over \alpha a}
\right)^n ,
\eqno (AP3-4)$$

\par\noindent
or

\boxit{
$$
I_{mn} =
\sqrt {{2 \pi k  \over a }}
\left(
{2k \over \alpha }\right)^{n/2}
\exp \left[
- {k \over 2 \alpha}
\left(
{2 \pi m  \over \alpha a}
\right)^2
\right]\ 
\left(
{2 \pi i m  \over \alpha a}
\right)^n .
\eqno (AP3-5)$$
}

\shead{APPENDIX 4}

\par
By definition

$$
J_{mn} = 
\int_{{-} \infty}^{\infty}
 \phi_{mn} v\  dv .
\eqno (AP4-1)$$

\par
According to definition of Fourier transformation we could express the value of first moment of any function through the first derivative value of its transform in zero point:

$$
J_{mn} = 
2 \pi i
F '_{mn} (0)
=
2 \pi i
F_{mn} (0)\ 
\left[
- {k \over \alpha} 
\left( 
q + { { 2 \pi m }  \over { \alpha a} }
\right) + 
{ n \over 
{\left( 
q - { { 2 \pi m }  \over {\alpha a} }
\right) } }
\right]_{q=0}
 .
\eqno (AP4-2)$$

\par
We have finally:

\boxit{
$$
J_{mn} =
\sqrt {{2 \pi k  \over a }}
\left(
{2k \over \alpha }\right)^{n/2}
\exp \left[
- {k \over 2 \alpha}
\left(
{2 \pi m  \over \alpha a}
\right)^2
\right]\ 
\left[
-
{k \over \alpha} 
\left( 
{ 2 \pi i m   \over \alpha a}
\right)^{n+1} + 
n\ 
\left( 
{ 2 \pi i m   \over \alpha a}
\right)^{n-1}
\right]
.
\eqno (AP4-3)$$
}
%
\end{document}